\documentstyle[12pt,epsf]{article}
\begin{document}
\hoffset=-1.2cm
\hsize=15.3cm
\vsize=25.5cm
\vskip 2cm
\noindent
{\bf {CHIRAL SYMMETRY BREAKING FOR FUNDAMENTAL FERMIONS}}
\vskip 1.5cm
\baselineskip=7.5mm
\hspace{2.54cm} A. Bashir
\vskip 5mm
\hspace{2.54cm} Centre for Particle Theory, 

\hspace{2.54cm} University of Durham,

\hspace{2.54cm} Durham DH1 3LE, U.K.
\vskip 1.5cm
\noindent
{\bf {INTRODUCTION}}
\vskip 2mm
Massive fermions have long been a problem in gauge theories.
Unification of electromagnetic and weak forces was once hindered by
the fact that the introduction of mass terms broke the gauge
invariance of the theory. This problem was solved by the introduction
of the
Higgs field. Spontaneous breakdown of the SU(2)
$\times$ U(1) symmetry then takes place. The gauge bosons gain mass
and the masses for the fermions are 
generated through their Yukawa interaction with this Higgs field. However,
there has been a widespread dissatisfaction with this mechanism since
the masses are not predictable. Rather, they must be fixed by
experiment. 
Studying the non-perturbative behaviour of gauge theories provides an
alternative. If the interactions are strong enough, they are capable
of generating masses for the particles 
dynamically even if they start with zero bare mass. Moreover,
experiment tells us that the top quark is
very heavy and so the Yukawa coupling $g_t$ for top-Higgs
interaction is O(1). Then one naturally expects that non-perturbative
effects become important. Indeed, it has been suggested
$[1]$ that the top quark 
may acquire mass non-perturbatively through four-fermion interactions, and
the Higgs can then be viewed as the condensate of the top and
the antitop. However, in an attempt to
include the effects of gauge boson exchange term, one
loses gauge invariance of the physical
quantities. Of course, physical quantities must be gauge independent.
This motivates the study of how to achieve this in non-perturbative
calculations. Quenched QED provides a toy model in which to study this
problem, as we discuss.  

\vskip 2cm
\noindent
{\bf{DYSON-SCHWINGER EQUATIONS}}
\baselineskip=7.5mm
\vskip 2mm
  Our starting point is the set of Dyson-Schwinger equations. These
are an infinite system of
coupled equations for all the Green's functions, which are
non-perturbative in nature. Their structure is such that the 1-point
function is related to the 2-point function, the 2-point function is
related to the 3-point function, etc. ad infinitum. As it is impossible to
solve the complete set of equations, one has to truncate this infinite
tower in a physically acceptable way to reduce them to something that is
soluble. A familiar way to do this is perturbation theory.
However, if one wishes to generate masses for particles, a
non-perturbative way has to be sought. 

To see how to do this consider two of the Dyson-Schwinger equations,
one for the fermion propagator, and the other for the photon
propagator. These are shown below diagrammatically together with their
corresponding mathematical expressions:   

\epsfbox[60 50 200 180]{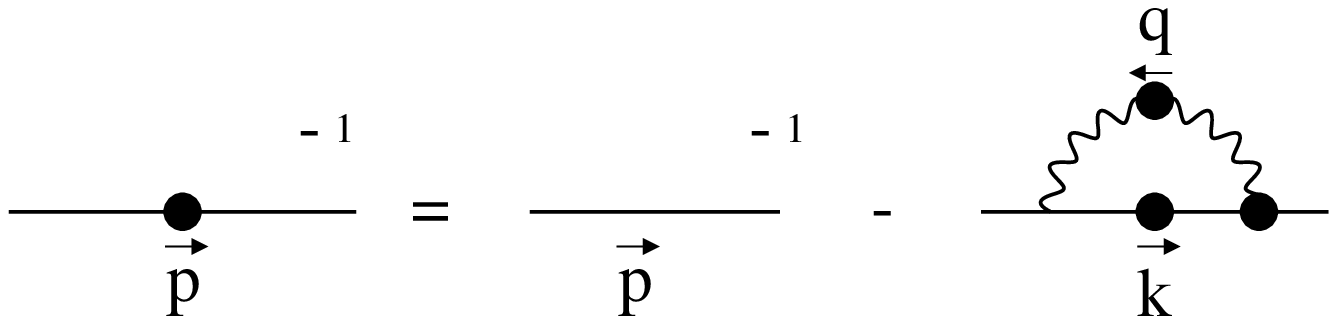}

\begin{small}
{\centerline{FIG. 1. Dyson-Schwinger equation for fermion
propagator.}}
\end{small}
\begin{eqnarray}
     iS_{F}^{-1}(p)\,=\,iS_{F}^{0^{-1}}(p)\,-\,e^2\int
         \frac{d^4k}{(2\pi)^4}\, 
    \gamma^{\mu}\,S_{F}(k)\, \Gamma^{\nu}(k,p)\,\Delta_{\mu\nu}
    (q) \quad ,  
\end{eqnarray}

\epsfbox[60 50 200 160]{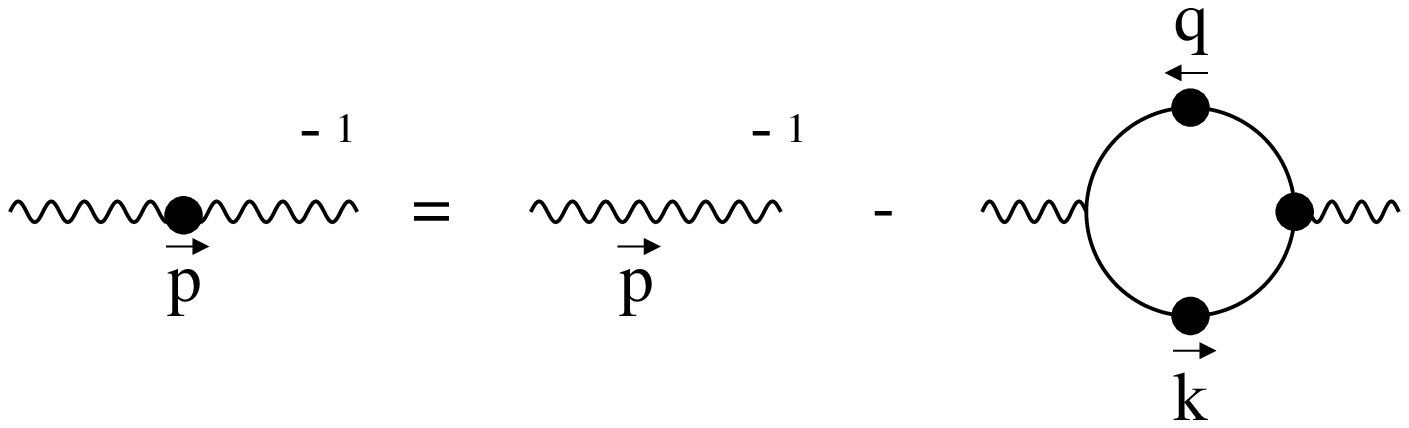}

\vskip 4mm
\begin{small}
{\centerline{FIG. 2. Dyson-Schwinger equation for photon propagator.}}
\end{small}
\begin{eqnarray}
     i\Delta_{\mu\nu}^{-1}(p)\,=\,i\Delta_{\mu\nu}^{0^{-1}}(p)\,-\,e^2N_f\int
         \frac{d^4k}{(2\pi)^4}\, 
    \gamma^{\mu}\,S_{F}(k)\, \Gamma^{\nu}(k,p)\,S_{F}
    (q) \quad ,
\end{eqnarray}


\noindent
where the quantities with the superscript '0' are bare
quantities, and the others are full ones. Quenched
QED corresponds to making the assumption that the full photon
propagator can be replaced by its bare counterpart. This limit is
achieved by regarding $N_{f}$ as a mathematical parameter, which is
set equal to zero. As an example, to begin with, we make a further
simplification by replacing 
the full vertex by the bare one. Eq. (1) then reduces to:

\epsfbox[60 50 200 180]{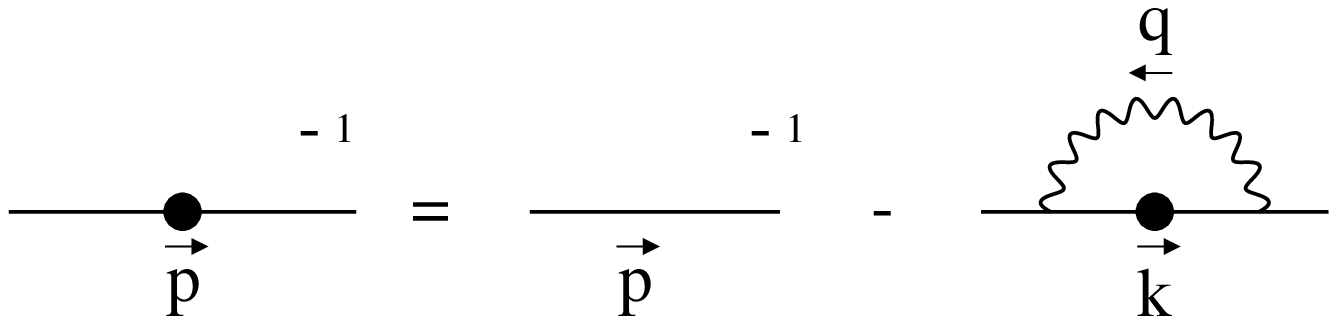}

\begin{small}
{\centerline{FIG. 3. Rainbow approximation.}}
\end{small}
\begin{eqnarray}
     iS_{F}^{-1}(p)\,=\,iS_{F}^{0^{-1}}(p)\,-\,e^2\int
         \frac{d^4k}{(2\pi)^4} \, 
    \gamma^{\mu}\,S_{F}(k)\, \gamma^{\nu}(k,p)\,\Delta_{\mu\nu}^0
    (q) \quad , 
\end{eqnarray}
in what is known as the rainbow approximation, where 
\begin{eqnarray*}
    S_{F}(k)&=& \frac{F(k^2)}{\not \! k- {\cal M}(k^2)}\ \quad , \\ 
    S_{F}^{0}(k)&=&\frac{1}{\not \! k - m_{0}}\    \quad ,     \\ 
    \Delta_{\mu\nu}^0(q)&=&\frac{1}{q^2}\ \left(
    g_{\mu\nu}+(\xi-1)\frac{q_{\mu}q_{\nu}}{q^2}\ \ \right) \quad .   
\end{eqnarray*}
 Eq. (3) is a matrix equation which corresponds to two equations in
${\cal M}$ and F. We can project out equations for these by taking the
trace of Eq. (3) having multiplied by $\not \! p$ and 1 in turn to
obtain:
\begin{eqnarray*}
\nonumber 
 \frac{1}{F(p^2)}\,&=&\; 1\;- \frac{\alpha}{4\pi^3} \frac{1}{p^2} \int
                          d^4k \,
                          \frac{F(k^2)}{k^2+{\cal M}^2(k^2)}\
                          \frac{1}{q^2}\  \cdot    \\  \nonumber   
                  && \hspace{20mm}  \Bigg\{ -2k\cdot  p   
                     - \frac{(\xi-1)}{q^2}\ \left[ 2
                       k^2 p^2 - (k^2 + p^2) k\cdot p
                       \right]    \Bigg\} \ ,  \\  \nonumber   
 \frac{{\cal M}(p^2)}{F(p^2)}\ &=& \; m_0\; -\frac{\alpha}{4\pi^3} \int 
                          d^4k \,
                          \frac{F(k^2){\cal M}(k^2)}{k^2+{\cal M}^2(k^2)}\ 
                          \frac{1}{q^2}\ (3+\xi) \ .
\end{eqnarray*} 
where as usual $\alpha = e^2/4\pi$.
\noindent
 On carrying out the angular integrations, and putting the bare mass
equal to zero, we have
\begin{eqnarray}
 \frac{1}{F(p^2)}\ &=& 1\, +\frac{\alpha\xi}{4\pi}\
                          \int_{0}^{\Lambda^2} 
                          dk^2\,  \frac{F(k^2)}{k^2+{\cal
                      M}^2(k^2)}\, 
                   \left[ \frac{k^4}{p^4} \theta (p^2-k^2) 
                          +  \theta (k^2-p^2) \right] \quad ,\\ 
 \frac{{\cal M}(p^2)}{F(p^2)}\ &=& \frac{\alpha(3+\xi)}{4\pi}\
                          \int_{0}^{\Lambda^2} 
                          dk^2\, \frac{{\cal M}(k^2) F(k^2)}{k^2+{\cal
                         M}^2(k^2)}\, 
                           \left[ \frac{k^2}{p^2} \theta (p^2-k^2)
                          +  \theta (k^2-p^2) \right] \ ,
\end{eqnarray}
where $\Lambda$ is the ultra-violet momentum cutoff. It is easiest to solve these equations in the Landau gauge where they
decouple. $F(p^2)$ is obviously $1$. Moreover, there is a non-trivial
solution $[2]$ for the mass function ${\cal M}$ for the coupling larger than
a critical value of $\alpha_{c}= \pi/3$. This is best illustrated by
plotting the Euclidean mass $M={\cal M}(M^2)$ as a function of
$\alpha$, as found by Curtis and Pennington $[3]$: 
\vskip 7.5cm

\epsfbox[80 60 200 180]{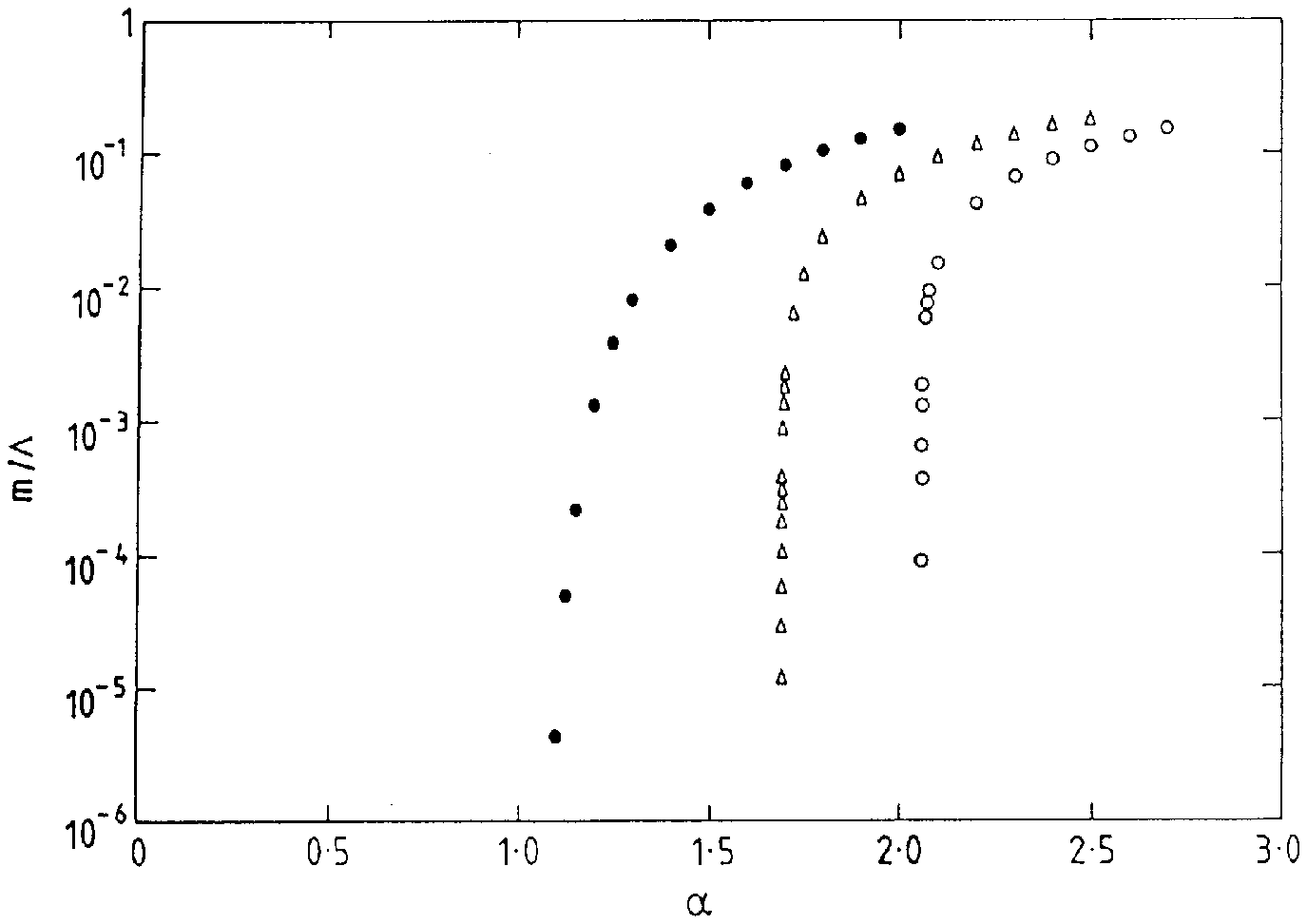}

\noindent
\baselineskip=5.5mm
\begin{small}
\baselineskip=5.5mm
FIG. 4. Euclidean mass, $M={\cal M}(M^2)$ dynamically
generated in the rainbow approximation as a function of the
coupling $\alpha$ in three different gauges: Landau $( \xi = 0 )$ $\bullet$,
Feynman $( \xi = 1) $  $\tiny \triangle$, and Yennie $( \xi = 3 )$  $\circ$ gauges.
\end{small}
\baselineskip=5.5mm
\vskip 3mm
\baselineskip=7.5mm
Note that ${\cal M}=0$ is always a solution to Eq.
(5). However, 
beyond the critical value of the coupling, the non-zero solution
bifurcates away from the trivial solution. 
This is in complete
contrast with the perturbation theory, where, even if we perform an all
orders resummation using the Renormalization Group Equation, we end up
with a result of the following form,
\begin{eqnarray*}
  {\cal M}(p^2)&=&m_0 X(p^2)   \\ 
  X(p^2)&=&\sum_{n}^\infty  \sum_{m}^n \alpha^{n} A_n B_{m,n}
     \ln^m(p^2/\Lambda^2)  \,
\end{eqnarray*}
and the field remains massless to all orders if we start with a
zero bare mass, $m_{0}=0$.

 In contrast, non-perturbative dynamics is able to generate
masses for particles even if they have zero bare mass. However,
there are problems. As the critical coupling corresponds to a change
of phase, we expect it to be independent of the gauge parameter. But
when one solves the Eqs. (4) and (5) for different gauges, one finds
that this is not the case, as depicted in Fig. 4. 
However, it is not
difficult to trace the root of this problem. The full vertex of Eq.
(1) has
to satisfy the Ward-Takahashi identity for 
the fermion propagator to ensure its gauge covariance. However, the
bare vertex that was used in Eq. (3) does not obey this identity.
Therefore, one should not expect physical outputs to be gauge
independent when the input is not.  
\vskip 2mm
\noindent
{\bf {THE VERTEX}}
\vskip 2mm
We expect that any reasonable {\em ansatz} for the vertex should
fulfill the following requirements:
\begin{itemize}
\item
    It must satisfy the Ward-Takahashi Identity in all gauges.
\begin{eqnarray*}
 q^{\mu} \Gamma_{\mu} = S_{F}^{-1}(k) - S_{F}^{-1}(p)
\end{eqnarray*}
\item
    It must ensure that the fermion propagator of Eq. (1) is
multiplicatively renormalizable.  
\item  
    It must result in a critical coupling, at which mass is generated
dynamically, that is gauge independent.
\item  
    It must be free of any kinematic singularities, i.e. it should
have a unique limit when  $k^{2} \rightarrow  p^{2}$. 
\item 
    It must have the same transformation properties as the bare vertex
$\gamma^{\mu}$ under $C$ and $P$.
\end{itemize}
Keeping in mind the form of the Ward-Takahashi identity, one can split
the full vertex into two components, longitudinal and transverse:
\begin{eqnarray}
      \Gamma^{\mu}(k,p)&=&\Gamma^{\mu}_{L}(k,p)+\Gamma^{\mu}_{T}(k,p)
\ ,
\end{eqnarray}
where, the transverse part of the vertex is defined by:
\begin{eqnarray}
        q_{\mu} \Gamma^{\mu}_{T}(k,p)&=&0  \ .
\end{eqnarray}
The Ward-Takahashi identity uniquely fixes the
longitudinal part of the vertex, as shown by Ball and Chiu $[4]$, to be
\begin{eqnarray} 
\Gamma^{\mu}_{L}(k,p)&=& a(k^2,p^2) \gamma^{\mu}
                       + \mbox{} b(k^2,p^2) (\not \! k + \not \! p)
                           (k+p)^{\mu}   \\    \nonumber
                      &-& \mbox{} c(k^2,p^2) (k+p)^{\mu} \ ,  
\end{eqnarray}
where
\begin{eqnarray}
\nonumber
  a(k^2,p^2)&=&\frac{1}{2}\ \left( \frac{1}{F(k^2)}\ +
  \frac{1}{F(p^2)}\  \right) \qquad  \quad \qquad,  \\  \nonumber     
  b(k^2,p^2)&=&\frac{1}{2}\ \left(
  \frac{1}{F(k^2)} - \frac{1}{F(p^2)}\ \right) \frac{1}{k^2 - p^2} \qquad , \\
       \nonumber  
  c(k^2,p^2)&=&  \left( \frac{{\cal M}(k^2)}{F(k^2)}\
  - \frac{{\cal M}(p^2)}{F(p^2)}\ \right) \frac{1}{k^2 - p^2} \qquad .   
\end{eqnarray}
 However, the transverse part remains arbitrary. Ball and Chiu $[4]$
enumerated a basis of eight independent tensors 
in terms of which the most general form for the transverse part of the
vertex can be written:
\begin{eqnarray}
   \Gamma^{\mu}_{T}(k,p) =  \sum_{i=1}^8 \tau_{i}(k^2,p^2,q^2)
   T_{i}^{\mu}(k,p)  \ .
\end{eqnarray} 
We list here only those four tensors which we shall need later:
\begin{eqnarray}
\nonumber
   T_{2}^{\mu}(k,p)&=& \left(p^{\mu}(k.q)-k^{\mu}(p.q) \right) ( \not
\! k +\not \! p) \\ \nonumber 
   T_{3}^{\mu}(k,p)&=& q^{2} \gamma^{\mu}-q^{\mu} \not \! q  \\ \nonumber
   T_{6}^{\mu}(k,p)&=& \gamma^{\mu} (k^{2} - p^{2})
                         -( k+p)^{\mu} ( \not \! k -\not \! p ) \\ 
   T_{8}^{\mu}(k,p)&=& -\gamma^{\mu} p^{\nu} k^{\rho} \sigma_{\nu\rho}
                       + p^{\mu} \not \! k - k^{\mu} \not \! p \ \ ,
\end{eqnarray}
with $\sigma_{\mu\nu}= \frac{1}{2} [\gamma_{\mu},\gamma_{\nu}]$. 
The simplest choice is to take the transverse part to be zero. But
Curtis and Pennington $[5]$ showed that if we 
take the transverse part of the vertex to be zero, the fermion
propagator is no longer multiplicatively renormalizable. They
suggested the following transverse part of the vertex
satisfying this requirement.
\begin{eqnarray}
   \Gamma^{\mu}_{T}(k,p) = \frac{1}{2}  \left(
    \frac{1}{F(k^2)}\ - \frac{1}{F(p^2)}\ \right) \frac{1}{d(k^2,p^2)}  
   T_{6}^{\mu}(k,p)  \ , 
\end{eqnarray}
where, $d(k^2,p^2)=k^{2}$ for $k^{2} \gg p^{2}$. $d(k^2,p^2)$ must be
symmetric in $k$ and $p$ and free of kinematic singularities leading
to the proposal: 
\begin{eqnarray}
   d(k^2,p^2)=  \frac{(k^2-p^2)^2+[{\cal M}^2(k^2)+{\cal
                 M}^2(p^2)]^2}{k^2+p^2}\  \ . 
\end{eqnarray}
The vertex specified by Eqs. (8-12) will be referred to as the
CP-vertex [5]. Curtis and Pennington solved the coupled equations for $F$ and ${\cal
M}$ from Eq. (1), using this {\em ansatz}. They found that the
gauge-dependence of the 
critical coupling at which the non-perturbative behaviour bifurcates away from
the perturbative one reduces considerably, as seen by comparing Figs.
4 and 5.
\vskip 7.5cm

\epsfbox[80 70 200 180]{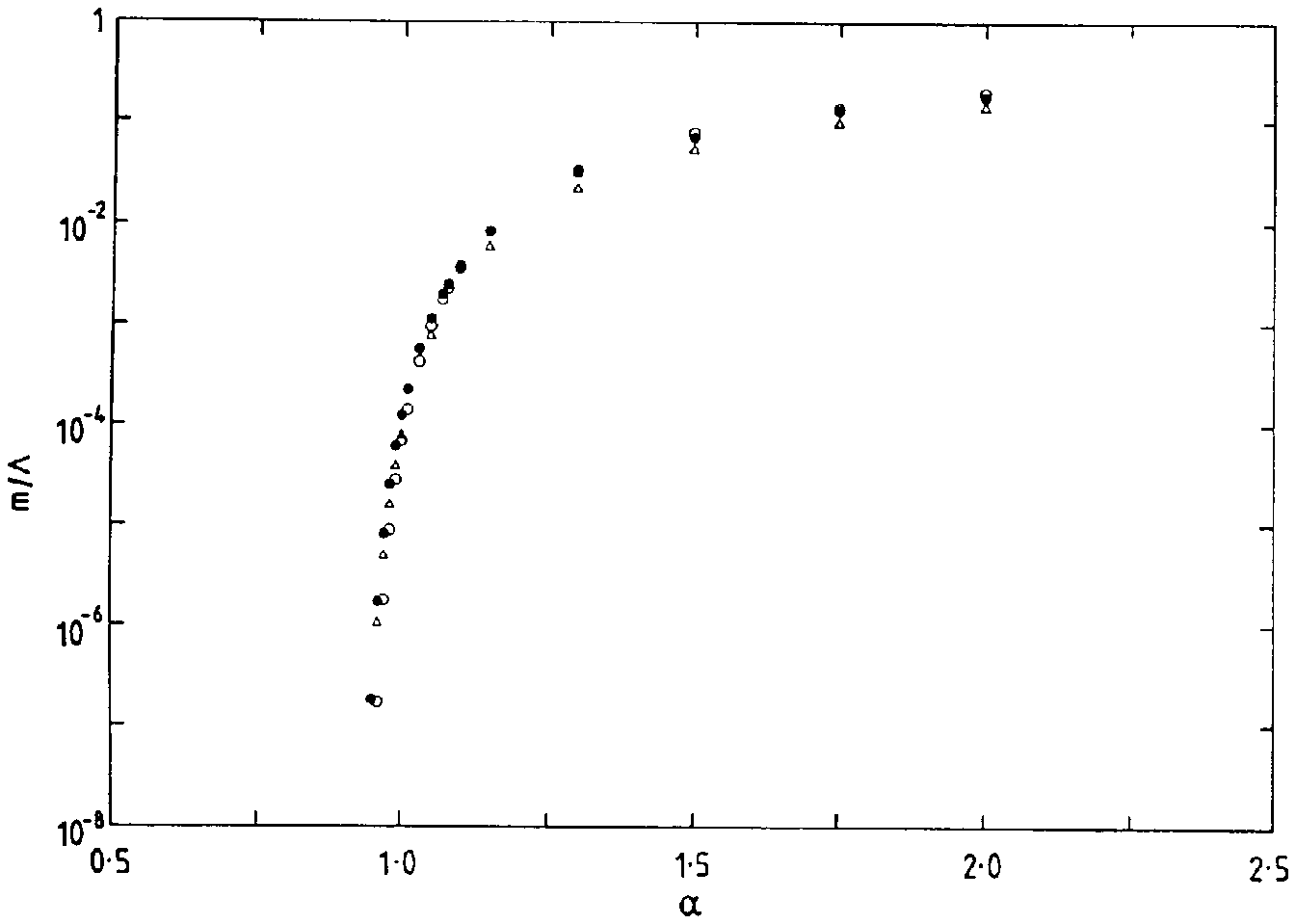}

\noindent
\baselineskip=5.5mm
\begin{small}
\baselineskip=5.5mm
FIG. 5.
 Euclidean mass, $M={\cal M}(M^2)$ dynamically
generated using the CP-vertex as a function of the
coupling $\alpha$ in three different gauges: Landau $( \xi = 0 )$ $\bullet$,
Feynman $( \xi = 1) $  $\tiny \triangle$, and Yennie $( \xi = 3 )$
$\circ$ gauges. This plot is to be compared with the rainbow
approximation results of Fig. 4.
\end{small}
\baselineskip=5.5mm
\vskip 3mm
\baselineskip=7.5mm
\noindent
\vfil\eject
{\bf{BIFURCATION ANALYSIS}}
\vskip 2mm
 To see this, Atkinson et al. $[6]$ recently suggested a bifurcation
analysis to study 
the phase change near the critical coupling. This is a precise
way to locate the critical coupling as compared to the previous
methods which rely on numerical calculations. 
This method amounts, in practice, simply to throwing away all terms that are
quadratic or higher in the mass-function ${\cal M}$. Employing this
procedure, and using the fact that at the critical coupling,
${\cal M}(p^2) \sim (p^2)^{-s}$ and $ F(p^2) \sim (p^2)^{\nu}$ in Eq. (1),
one arrives at the following equation in an arbitrary gauge:
\begin{eqnarray*}
\nu &=& \frac{\alpha \xi}{4 \pi}       \\
\nonumber \xi &=& \frac{3 \nu (\nu- s+1)}{2 ( 1-s ) } \Bigg[ 3 -\pi \cot \pi(
       \nu-s ) + 2 \pi \cot \pi s - \pi \cot \pi \nu     \\
  &+& \frac{1}{\nu}\ +
     \frac{1}{\nu+1}\  + \frac{1}{\nu}\ +  \frac{2}{1-s}\ +
     \frac{3}{s- \nu}\ + \frac{1}{s- \nu -1} \Bigg] \quad .
\end{eqnarray*}
There are two roots of this latter equation for $s$ between 0 and 1. 
Bifurcation occurs when the two roots for $s$ merge at a point 
specified by $\partial \xi / \partial s = 0$ .
\vskip 7cm

\epsfbox[80 70 200 180]{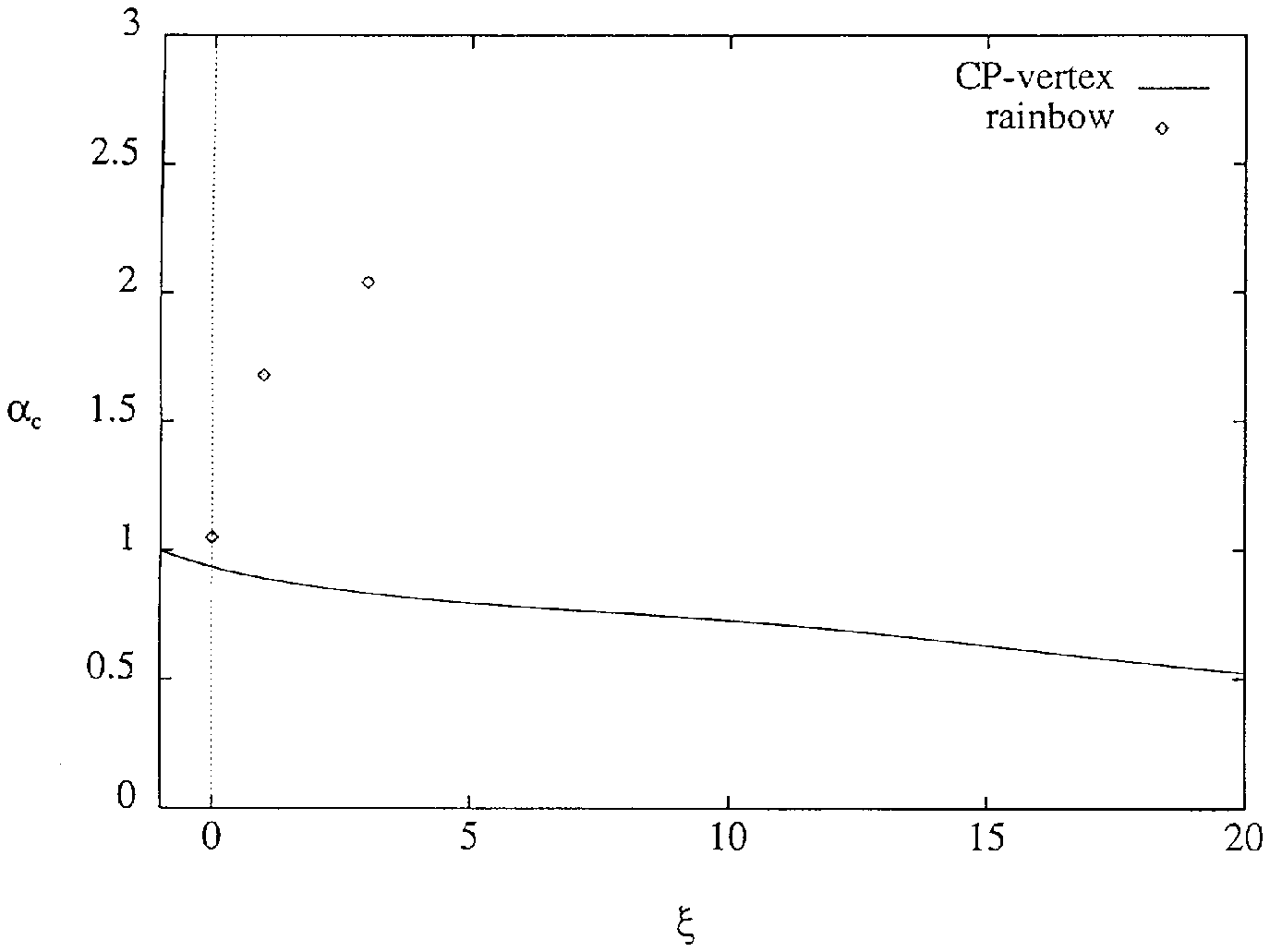}

\noindent
\baselineskip=5.5mm
\begin{small}
FIG. 6. Critical coupling, $\alpha_{c}$, as a function of the gauge
parameter, $\xi$ (solid line). The corresponding values for the
rainbow approximation  have also been shown $\diamond$.
\end{small}
\baselineskip=5.5mm
\vskip 3mm
\baselineskip=7.5mm
\noindent
The bifurcation point defines the critical coupling, $\alpha_{c}$.
Numerically, $\alpha_{c} = 0.933667$ in the Landau gauge.
For each
value of the gauge parameter, these equations can be solved for $\nu$,
$s_{c}$ 
and  $\alpha_{c}$. 
The solution found by Atkinson et al $[6]$ is displayed in
Fig. 6. For comparison, the points for the bare vertex
have also been shown. One can see that the gauge dependence has
considerably been reduced, as was seen earlier. However weak
this variation, any gauge dependence shows that the {\em CP} vertex
cannot be the exact choice. 
\vskip 2mm
\noindent
{\bf {CONSTRAINTS OF MULTIPLICATIVE RENORMALIZABILITY}}
\vskip 2mm
 To find a vertex that ensures the gauge independence of the critical
 coupling, we start off by making three assumptions.
 Firstly, we demand that a chirally-symmetric solution 
 should be possible when the bare mass is zero, just as in
  perturbation theory. 
  This is most easily accomplished if
  the sum in Eq. (9) involves just $i=2,3,6$ and 8.
  The second assumption is  that the functions, $\tau_{i}$, 
  multiplying the transverse
  vectors, Eq. (9), only depend on $k^2$ and $p^2$, but {\bf not} $q^2$. The
third assumption is that the transverse part of the vertex vanishes in
the Landau gauge. The motivation for this comes from the lowest order
perturbative calculation for the transverse vertex, satisfied by Eq.
(11). These conditions fix the $\tau_{i}$ of Eq. (9). Multiplicative
renormalizability of the wavefunction renormalization $F(p^{2})$
enables us to write $\tau_{6}$ and $\overline{\tau}$ in terms of one
function $W_{1}(x)$ $[7]$ :   
\begin{eqnarray}
   \overline{\tau}(k^2,p^2)\;= && \;\ \frac{1}{4}\ \frac{1}{k^2-p^2}\
        \frac{1}{s_{1}(k^2,p^2)}\  \Bigg[
         W_{1}\left( \frac{k^2}{p^2}\ \right) - W_{1}\left(\frac{p^2}{k^2}\
         \right) \Bigg]  \ , \\   \nonumber
\tau_{6}(k^2,p^2)\;=&&-\,\frac{1}{2}\ \frac{k^2+p^2}{(k^2-p^2)^2}
    \left( \frac{1}{F(k^2)} - \frac{1}{F(p^2)} \right)      +
      \frac{1}{3}\, \frac{k^2+p^2}{k^2-p^2}\ \overline{\tau}(k^2,p^2) \\
  &&+ \,\frac{1}{6}\ \frac{1}{k^2-p^2}\  \frac{1}{s_{1}(k^2,p^2)}\
         \Bigg[ W_{1}\left(\frac{k^2}{p^2}\ \right)  +
        W_{1}\left(\frac{p^2}{k^2}\ \right)  \Bigg]  \ ,      
\end{eqnarray}
where $\overline{\tau}$ is the combination of $\tau_i$ given by:
\begin{eqnarray*}
  \overline{\tau}(k^2,p^2)&=&\tau_{3}(k^2,p^2) + \tau_{8}(k^2,p^2) -
                  \frac{1}{2}\ (k^2+p^2)\, \tau_{2}(k^2,p^2) \ ,
\end{eqnarray*}
and
\begin{eqnarray*}
   s_{1}(k^2,p^2)&=& \frac{k^2}{p^2}\ F(k^2) +  \frac{p^2}{k^2}\
F(p^2)  \ . 
\end{eqnarray*}
 The condition of multiplicative renormalizability, i.e,  $ F(p^2)
\sim (p^2)^{\nu}$, constrains the otherwise arbitrary function $W_1$
as follows: 
\begin{eqnarray*}
       \int_{0}^{1}  dx \;W_1(x)=0  \ .
\end{eqnarray*}
 It should be noted that, with the simplest choice  $W_1=0$, the
massless CP-vertex, Eqs. (11,12) emerges.  
\vskip 2mm
\noindent
{\bf {CONSTRAINTS OF GAUGE INVARIANCE}}
\vskip 2mm

At the bifurcation point, as stated before, multiplicative renormalizability 
forces a simple power behaviour for the mass function as well as
for the wavefunction renormalization. Such a multiplicatively
renormalizable mass function must exist in all gauges. Consequently,
the exponent, $s_c$, must be gauge independent.  Moreover, dynamical mass
generation marks a physical phase change and so the critical coupling,
$\alpha_c$, must also be gauge independent. Thus
the critical values, $\alpha_{c}$,  $s_{c}$, found in the
Landau gauge must hold in all gauges. Using this physically motivated
argument, the equation for the mass function gives $\tau_{2}, \tau_{3}$
and $\tau_{8}$ in terms of a function $W_2(x)$ $[8]$:  

\begin{eqnarray*}
         \tau_2(k^2,p^2)&=& \frac{2\xi}{(k^2-p^2)^2} \,\frac{
          q_2(k^2,p^2)}{ s_2(k^2,p^2)} -6 \,\frac{ \tau_6(k^2,p^2)}{
          (k^2-p^2)} \\ 
       & &-\frac{1}{(k^2-p^2)^2}\  \frac{1}{s_2(k^2,p^2)}\
           \Bigg[ W_2\left( \frac{k^2}{p^2}\ \right) +  W_2\left(
            \frac{p^2}{k^2}\ \right) \Bigg]  \\ 
       & &-\frac{k^2+p^2}{(k^2-p^2)^3}\  \frac{1}{s_2(k^2,p^2)}\
           \Bigg[ W_2\left( \frac{k^2}{p^2}\ \right) -  W_2\left(
            \frac{p^2}{k^2}\ \right) \Bigg] \quad ,   \\ 
          \tau_3(k^2,p^2)&=&  - \frac{k^2+p^2}{
          k^2-p^2}\, \tau_6(k^2,p^2) \\
          & &+ \frac{1}{k^2-p^2}\ 
          \frac{1}{s_2(k^2,p^2)}\ \Bigg[ \frac{1}{2}\ r_2\left(
          \frac{k^2}{p^2}\ \right) - \frac{\xi}{3} 
          \,q_3(k^2,p^2) \Bigg]    \\ 
          & &- \frac{1}{6} \frac{k^2+p^2}{(k^2-p^2)^2}\ 
          \frac{1}{s_2(k^2,p^2)}\  \Bigg[ W_2\left( \frac{k^2}{p^2}\ \right) +
          W_2\left( \frac{p^2}{k^2}\ \right) \Bigg]  \\ 
          & & +\frac{1}{6} \frac{k^4+p^4-6k^2p^2}{(k^2-p^2)^3}\ 
          \frac{1}{s_2(k^2,p^2)}\  \Bigg[ W_2\left( \frac{k^2}{p^2}\ \right) -
          W_2\left( \frac{p^2}{k^2}\ \right) \Bigg] \quad ,   
\end{eqnarray*}
and
\vfil\eject
\begin{eqnarray*}
          \tau_8(k^2,p^2)&=&  -2\, \frac{k^2+p^2}{
          k^2-p^2} \,\tau_6(k^2,p^2) + \overline{\tau}(k^2,p^2) \\ 
          & &-  \frac{1}{k^2-p^2}\ 
          \frac{1}{s_2(k^2,p^2)}\ \Bigg[ \frac{1}{2}\ r_2\left(
          \frac{k^2}{p^2}\ \right) - \frac{\xi}{3}
           \,q_8(k^2,p^2)     
          \Bigg]        \\ 
          & &- \frac{1}{3}\ \frac{k^2+p^2}{(k^2-p^2)^2}\
            \frac{1}{s_2(k^2,p^2)}\   
           \Bigg[ W_2\left( \frac{k^2}{p^2}\ \right) +  W_2\left(
            \frac{p^2}{k^2}\ \right) \Bigg]   \\ 
          & &- \frac{2}{3}\ \frac{k^4+p^4}{(k^2-p^2)^3}\
            \frac{1}{s_2(k^2,p^2)}\   
           \Bigg[ W_2\left( \frac{k^2}{p^2}\ \right) -  W_2\left(
            \frac{p^2}{k^2}\ \right) \Bigg] \quad ,  
\end{eqnarray*}
where
\begin{eqnarray*}
            r_1\left(\frac{k^2}{p^2}\right)&=&
            \left(\frac{k^2}{p^2}\right) \left[1-
            \left(\frac{k^2}{p^2}\right)^{\nu}\right]
            -\left(\frac{k^2}{p^2}\right)^{-1}
            \left[1-\left(\frac{k^2}{p^2} \right)^{-\nu}\right] \quad , \\
           r_2\left(\frac{k^2}{p^2}\right)&=& \left(
            \frac{k^2}{p^2}\right)^{\frac{1}{2} -s_c}\ \left[1
            -\left(\frac{k^2}{p^2}\right)^{\nu}\right]  -
           \left(\frac{k^2}{p^2}\right)^{s_c - \frac{1}{2}\ }\
           \left[1-\left(\frac{k^2}{p^2}\right)^{-\nu}\right] \quad,  \\
 s_{1}(k^2,p^2)&=&\frac{k^2}{p^2}\ F(k^2) +  \frac{p^2}{k^2}\ F(p^2)
\quad , \\
     s_2(k^2,p^2)&=&\frac{k}{p}\,  \frac{{\cal M}(k^2)}{{\cal M}(p^2)}\,
             F(k^2) + \frac{p}{k}\,  \frac{{\cal M}(p^2)}{{\cal
             M}(k^2)}\, F(p^2)  \quad ,   \\
  q_2(k^2,p^2)&=&\frac{1}{k^2-p^2} \Bigg[ \frac{k^3}{p}\,
              \frac{{\cal M}(k^2)F(k^2)}{{\cal M}(p^2)F(p^2)}
               -  \frac{p^3}{k}\,
              \frac{{\cal M}(p^2)F(p^2)}{{\cal M}(k^2)F(k^2)} \Bigg]
  \quad ,  \\
 q_3(k^2,p^2)&=&\frac{kp}{(k^2-p^2)^2} \Bigg[ (p^2-3k^2)\, 
               \frac{{\cal M}(k^2)F(k^2)}{{\cal M}(p^2)F(p^2)} - (k^2-3p^2)\,
               \frac{{\cal M}(k^2)F(k^2)}{{\cal M}(p^2)F(p^2)}
\Bigg]  \quad ,   \\
 q_8(k^2,p^2)&=&\frac{1}{(k^2-p^2)^2} \Bigg[\frac{k}{p}(3k^4+p^4)\, 
                \frac{{\cal M}(k^2)F(k^2)}{{\cal M}(p^2)F(p^2)} -
                \frac{p}{k}(k^4+3p^4)\, \frac{{\cal
                M}(k^2)F(k^2)}{{\cal M}(p^2)F(p^2)}  
                \Bigg]    \quad .
\end{eqnarray*}
and the function $W_2$ is constrained, by the gauge invariance of the
mass function and the critical coupling, to obey the following integral
equation, 
\begin{eqnarray*}
\int_{0}^{1}  {dx\over{\sqrt{x}}} \; W_2(x)\;\ =\;0\quad ,
\end{eqnarray*}
at the critical coupling $\alpha=\alpha_c$.
In order to make sure that none of the functions $\tau_{i}$ has
kinematic singularities as $k^{2} \rightarrow  p^{2}$, $W_1$ and  $W_2$
should also satisfy the following conditions: 
\begin{eqnarray*}\nonumber
   W_1(1) + W_1^{\prime}(1)\,&=&\;-6 \nu\; ,\quad \\
   W_2(1) + 2W_2^{\prime}(1)\,&=&\;2\xi(\nu-s-1)\quad .
\end{eqnarray*}
This defines the construction of the full vertex via Eqs.
(6-10) $[8]$. 
\vskip 2mm
\noindent
{\bf {CONCLUSIONS}}
\vskip 2mm
 Above we have presented a truncation of the fermion Schwinger-Dyson
equation for the quenched QED, 
 which  respects the key properties of the theory. 
 We have constructed a non-perturbative vertex in terms of the
constrained functions $W_{i}(x) (i=1,2)$. It satisfies
the Ward-Takahashi identity,  
 ensures the fermion
 propagator is multiplicatively renormalizable,
  agrees with one loop perturbation
 theory for large momenta and enforces a gauge independent chiral symmetry
 breaking phase transition. 
  This study motivates the need for a realistic investigation of
 $t{\overline t}$ condensates as the source of the electroweak
 symmetry breaking. Including the four fermion interaction, the
Dyson-Schwinger equation for the fermion propagator becomes:
  
\epsfbox[75 50 200 180]{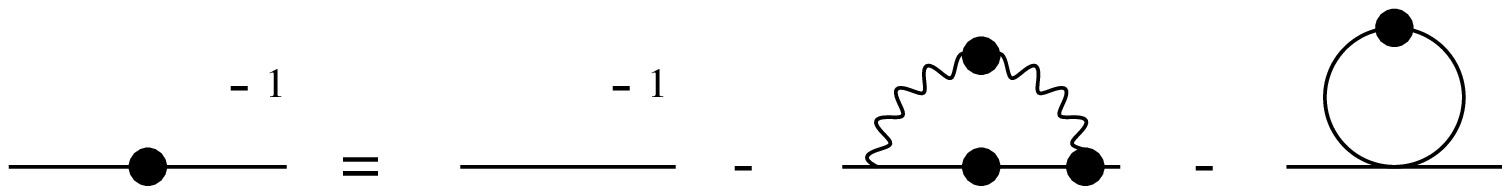}

\noindent
\baselineskip=5.5mm
\begin{small}
FIG. 7. Dyson-Schwinger equation for the fermion propagator, including
the four fermion interaction term. 
\end{small}
\baselineskip=5.5mm
\vskip 2mm
\baselineskip=7.5mm
We need to solve this equation in a gauge invariant way. The study
of quenched QED presented here suggests that a proper choice of the
vertex can guarantee the gauge independence of the physical observables.
However, a realistic calculation, of course, requires the unquenching
of the theory which complicates the problem significantly. The
fermion-boson vertex (in particular its transverse part) will 
intimately depend on the photon renormalization function in a
non-perturbative way not yet understood. The discussion for quenched QED
presented here provides the 
starting point for such an investigation of full QED.
\vskip 2mm
\noindent{\bf ACKNOWLEDGEMENTS}
\vskip 2mm
   This work was performed in collaboration with  M.R. Pennington. 
   I wish to thank the Government of Pakistan for a research
   studentship and the University of Durham and Institut d'Etudes
   Scientfiques de Carg\`{e}se for providing me with the funds to
   attend the School. 

\vskip 2mm
\noindent{\bf REFERENCES}
\vskip 2mm
\noindent
[1] W.A. Bardeen, C.T. Hill and M. Lindner, Phys. Rev.
{\bf D41} 1647 (1990).
\vskip 2mm 
\noindent
[2] V.A. Miransky, Nuovo Cim. {\bf 90A} 149 (1985)~;\\
\indent Sov. Phys. JETP
{\bf 61} 905 (1985)~;\\
\indent P.I. Fomin, V.P. Gusynin, V.A. Miransky and Yu.A. Sitenko, \\
\indent La rivista del Nuovo Cim. {\bf 6}, numero 5, 1 (1983).
\vskip 2mm 
\noindent
[3] D.C. Curtis and M.R. Pennington, Phys. Rev. {\bf D48}
4933 (1993). 
\vskip 2mm 
\noindent
[4] J.S. Ball and T.W. Chiu, Phys. Rev. {\bf D22} 2542 (1980).
\vskip 2mm
\noindent
[5] D.C. Curtis and M.R. Pennington, Phys. Rev. {\bf D42}
4165 (1990).
\vskip 2mm 
\noindent
[6] D. Atkinson, J.C.R. Bloch, V.P. Gusynin, M.R.
Pennington and \\
\indent M. Reenders, Phys. Lett. {\bf B329} 117 (1994).
\vskip 2mm 
\noindent
[7] Z. Dong, H.J. Munczek and C.D. Roberts, preprint\\
\indent ANL-PHY-7711-94. 
\vskip 2mm
\noindent
[8] A. Bashir and M.Pennington, `` Gauge Independent Chiral
Symmetry \\
\indent Breaking in Quenched QED '', University of Durham preprint
DTP-94/48 \\ 
\indent(June, 94)  Phys. Rev. (to be published).


\end{document}